**Global Internet-based Crisis Communication: A case Study on SARS**

Goh Ong Sing
Kolej Universiti Kebangsaan Malaysia
goh@kutkm.edu.my

## Abstract

Crisis Communication is an effective communication mechanism in the world. The outbreak of the SARS disease and the way information on it was disseminated has illustrated the importance of effective and efficient crisis communication management. Consequently, we would like to highlight the viability of incorporating an intelligent agent software robot into a crisis communication portal (CCNet) to send the alert news to subscribed users via email and others mobile services such as SMS, MMS and GPRS. This study is in the framework design and prototyping stage. The proposed system consists of the integration of an intelligent agent, which comprises of a conversation agent that is employed to gain trust from the portal's users, and an Automated Knowledge Extraction Agent, which retrieves first hand information from sources such as WHO.org, ABCNews.com and Sars.gov.sg for public awareness.

## 1.0 Introduction

Smith (2003) highlighted the importance of corporations quickly making decisions during a crisis. She goes on explaining that the Internet has the capability of affecting every aspect of a crisis and thus if it is correctly utilized, it can be an indispensable aid to a public relations professional in the times of crises. The Life Services Network (1996) states that by doing a certain amount of planning in the form of "*what if?*" scenarios, the good of turning a potentially negative situation into a positive one is a real possibility. The recent outbreak of the Severe Acute Respiratory System (SARS) disease is something that needed to be handled carefully. Based on the data provided by the World Health

Organization (WHO,2004,2003) , as of May 12 2003, there have been a total of 7447 reported cases with over 552 fatalities . Until today, there is not any specific vaccination or possible cure. A more frightening fact is that the virus is very infective, can survive for a long period of time and is now mutating into several strains. This case study will be based on the SARS disease. There is a need for information to be disseminated to decrease communication breakdown during this crisis.

A new medium has appeared with the introduction of the Internet and it has increased the risks associated with information i.e. free accessibility, interactivity and globalization. Connectivity of personal, economical, political and media communication, on the other hand, have all led to a loss of journalistic control over the information market. For example, United States and Canadian regulators have warned more than 40 website operators to stop making unproven claims about preventing or treating SARS with items such as purifiers and herbs. Howard Beales, Director, Bureau of Consumer Protection of the United States Federal Trade Commission highlighted the fact that some Internet companies have been taking advantage of fears about SARS to promote a long list of herbal supplements ("US And Canada,"2003).

Many of the people managing public health crises are not crisis communication experts. Besides that, many of them do not have access to crisis communication experts and this leads to communication breakdown. In fact, they are not aware of a field known as crisis communication. As a result, they do not know the skills they needed and only do what they know best. For physicians, that may well be *ex cathedra* pronouncements about what people [1]thought to be feeling and doing. As for politicians, it may well be carefully crafted sound bites aimed at coming across as confident and in control (Halavais,2000) . Underlying these communication efforts, our contention that SARS crisis managers do need crisis communication counseling to manage Internet-based public health crises in the future.

In order to get a better grasp of crisis communication on the Internet, we will use empirical research methodology to analyze the consequences of modernity in terms of linking patterns of Internet-based crisis communication and trust. Trust is normally a means to compensate for our lack of knowledge in handling complex systems and trust reduces uncertainty. It does not make up for lack of knowledge, but it allows us to believe and act as if we were in a state of full and certain knowledge. If a construction is public and part of a common knowledge, this could be analyzed as a result of public communication.

---

[1] This report describes an ongoing research at the Centre for Artificial Intelligence and Modeling at the Faculty of Information and Communication Technology, KUTKM. Support for this work was received from the PJP / 2003 / FTMK (1) SO17 grant.

Crisis communication on a global level via the Internet works very clearly in the coverage and worldwide communication about the SARS disease. From a special point of view, the crisis communication of anthrax attacks in the September 11th events, the subsequent SARS and Bird Flu attacks, is a paradigm in crisis communication and especially in crisis communication where the Internet plays a crucial role. Therefore, we will use the cases on SARS to highlight a few aspects of the interrelationship between crisis communication and the Internet by deploying an intelligent agent called the software robot. This system can be used in the future for any crisis communication that might occur.

Despite the growth in information and communication technology, computerized application is required to have some basic intelligence. Moreover, Artificial Intelligence (AI) has gained its place in the information technology (IT) world and requires for a perfect humanized computing environment, which is present in this research. Undoubtedly, humanized computer application is the current trend in the IT world. In line with this trend, we propose to incorporate an intelligent agent software robot-Artificial Intelligence Neural-network Identity (AINI) and customized Artificial Intelligence Markup Language (AIML) servable knowledge base to serve as a real software robot in the CCNet Portal. This portal will also use latest technologies such as mobile chat and PDA chat, which in turn will be used to send text-based information and images on the SARS disease to subscribed users.

AINI can improve customer service by reducing customer reliance on live agent to provide better answers and information to service inquiries. In addition, this software robot incorporates personalization techniques by using technologies such as profile recognition, user configuration, click stream analysis, human touch interface, live interaction, multilingual knowledge base support, segmentation or rules based personalization, collaborative filtering and present a statistical reports for data analysis. This software robot also exploits natural human social tendencies to convey internationality through motor actions and facial expressions because one of the most important processes in the formation and maintenance of a relationship is that of self-disclosure. It is the act of revealing private and personal information to others.

## 2.0 Objectives

The objectives of this research are:

a.  To develop a crisis communication portal (CCNet portal) with the latest information for public awareness to tackle new diseases such as SARS, Bird Flu or other crisis communication. Rumors, hoaxes, false and bias information, which will affect informational uncertainty, will not to be considered in this development. This will offer users with a new alternative information disseminator in the Internet, which can be perceived as more trusted and credible.

b.  To bring about a new era by transforming traditional websites to the humanoid websites by deploying Artificial Intelligent Neural-network Identity (AINI), which is integrated with human-like avatar (doctor/nurse) 3D character, speech technology (Text-to-Speech) and software robot. In other words, users can interact and communicate naturally with the software robot at the website or via mobile services.

c.  To develop an automated Knowledge Extraction Agent to automatically build a knowledge base for AINI from the existing websites or online documents by using a crawler.

d.  To provide unlimited Internet communication between the users and the intelligent agent using web services technologies such as SMS, MMS or GPRS. This dimension constitutes the globalization of medium communication.

e.  To develop new network characteristics of the Internet which enable distribution to an unlimited audience and becomes multi-directional and suitable for intercultural communication, especially in Malaysia's multicultural society.

f.  To minimize communication breakdown during crisis communication because chaos normally sets in due to lack of information.

g.  To reduce the barrier to knowledge access because of the growth in information and communication. This growth has resulted in computerized application being required to have some basic intelligence. Moreover, Artificial Intelligence (AI) has gained its place in the information technology (IT) world nowadays and requires for a perfect humanized computing environment, which is now in this research.

h.  To improve customer service and provide better answers to service inquiries by employing an intelligent agent software robot, which exploits natural human social tendencies to convey attention through motor actions and facial expressions.

i.  To apply agent technology in intelligent interfaces in order to enhance the trust factor in web sites by humanizing the interface design factor.

j.  To increase the motivation of users by presenting an engaging manner, resulting in longer interaction times and higher quality of healthcare.

**3.0 Methodology**

Generally, there are two main components in the proposed system. They are the integration of Artificial Intelligent Neural-network Identity (AINI) engine into the CCNet portal, PDA application and mobile technologies. Besides that, the Knowledge Extraction Agent is also developed to serve as an automated knowledge base for AINI of the CCNet portal.

**3.1 CCNet Portal**

The research on the SARS disease needs detail collection of information on SARS from various sources like WHO, Ministry of Health, related websites on SARS, newspapers, papers, magazines etc. With the available resources, the development of the CCNet Portal that contains the latest information on SARS (statistics on the number of SARS cases will be integrated live from official resources) will be done using a crawler. This crawler will pick up news and content match to the given selection criteria from various websites. With the information at hand, the crawler will fetch them back to AINI's knowledge base and the URL database for future processing. As mentioned in the precious section, rumors, hoaxes, false and bias information, which will affect informational uncertainty will not be considered in this development. In this research, there are both **disembedded** and **embedded** interfaces with software robot which are used in a comparative study. Thus, the effectiveness of using an embedded software robot in the CCNet Portal will be proven. Other technologies such as SMS, GRPS, MMS and PDA will be integrated in the Front-End side of the CCNet Portal for user facilities. Users will be alerted of the latest updates in the CCNet Portal via these technologies.

### 3.2 Integration of Software Robot on the CCNet Portal

According to Clark (1999), the process of communication with computers may be seen as a form of disembodied language when these computers are regarded as agents and not as tools. In line with this, the proposed system will incorporate a new form of user interaction, which leverages natural language through the chatting process. FAQs related to SARS and other crisis communication, including the causes of the disease, prevention, medication, symptoms etc is going to be answered by the software robot. As for the backbone of the system, the AINI software robot that was developed by Goh Ong Sing will be used in this research. This Intelligent Agent knowledge-based robot will handle all the queries from the users using the Natural Language Processing (NLP) methodology, which can process AIML (Artificial Intelligent Markup Language). The Flash-MX technology was utilized to design a 3D human-like interface (doctor or nurse) for AINI. This feature will be embedded in the CCNet Portal. Besides that, an intelligent agent known as a Knowledge Extraction Agent will also be designed for providing information to the AINI knowledge base.

### Application of Mobile Technology on the CCNet Portal

Besides web-based application technologies, the building block of the current technologies such as PDA chat and mobile chat will be developed in this project. This service (the application of mobile technology) is provided to users upon subscription of the service from the CCNet portal. The questions placed by the users will be answered by AINI based on the powerful knowledge-based resources using Natural Language processing. Meanwhile, there will also be services such as SMS,

MMS and GPRS for the subscribed users. The AlertNews feature available on the CCNet Portal will alert users of any latest news. Thus, the incorporation of the latest technologies in the CCNet portal allows it to effectively handle crisis communication.

## 4.0 Architectural Overview

The architectural design of the system is shown in Figure 1.0. The Intelligent CCNet Portal can be divided into two main parts according to the operating environment, which are Front-End and Back-End .

The AINI Server and Mobile Gateway are located in the middle to function as the interconnection path between the Front-End (Client) and the Back-End (Server) of the system. It will process communication between the users of the system and the CCNet Portal. AINI's engine is a unique intelligent agent framework. All communication with AINI is done through a natural interface that uses a natural language processing technology and speech technology via a 3D animated character called avatar. AINI's engine implements its sophisticated decision making network based on the information it encounters in the knowledge bases. These decision-making capabilities use the XML specifications. It accepts questions and requests from users and processes the queries based on the information contained in AINI's knowledge base. The system architecture is shown below:

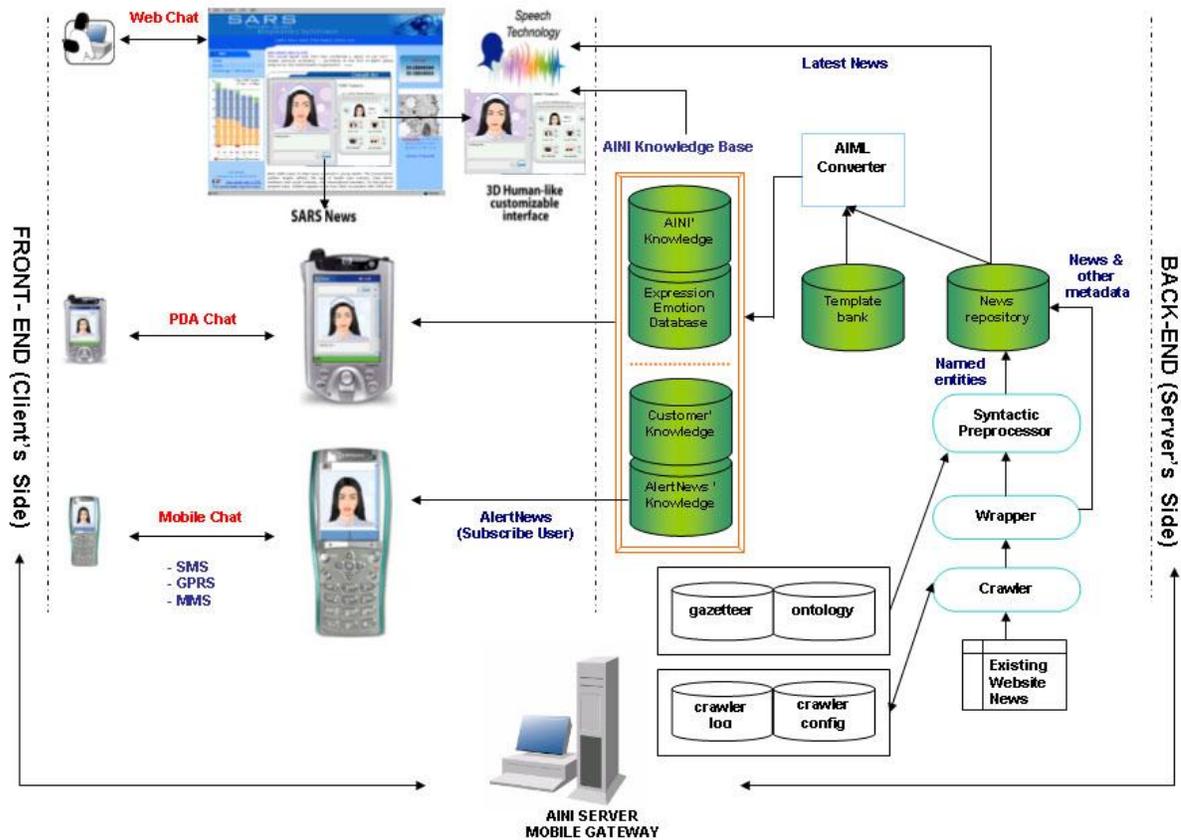

Figure 1.0: CCNet Portal

## 4.1 Front-End (Client's Side)

This section will provide emphasis on computer human interaction (HCI), which will involve users' interaction with the system. They can interact with AINI through three different methods, which are web chat, PDA chat and mobile chat. The following sections provide detailed descriptions of the CCNet Portal.

### 4.1.1 Web Chat

The web chat sessions allow the users to interact in real-time with the software robot at the website. These web sessions can either be text-based or voice-based with a 3D animated character and Text-to-Speech technology. Users are able to customize the interface, type questions and receive text responses directly from a website. Besides that, users can go through all the information on the website on the topics they are interested in. At the same time, they can place questions for more guidance on other topics. A collaborative browser allows a portal to guide the users through the organization's website by automatically "pushing" URLs and information from other websites to the user's browser. This not only facilitates communication between the software robot and users, but also allows the intelligence software robot to help users locate specific information on their websites. This is because AINI is able to intelligently react to interact with the user's commands.

AINI's engine which uses Artificial Intelligence and Natural Language Processing is an important targeted technology for the CCNet portal. By using a human-like software robot, users will have the feeling of interacting with another human being who will act on their command. So, the main objective of AINI is to intelligently offer related information on various topics (e.g. SARS), where the service requester is in a virtual environment where no real live agents or specialists will be physically involved. This means AINI will use natural language parsing i.e. AIML and the AINI engine to search the predefined knowledge base as well as other data sources located in different systems via networking. The virtual agent and the live agent compliment each other to enhance the interactive virtual advisor.

Users interact with the advisor through normal Internet ports, which are connected to AINI's knowledge base that provides WebGuide, WebTips and WebSearch engines. The purpose of WebGuide is to guide users through the entire portal. It will enable AINI to offer help without waiting for the surfer to ask. The WebTips engine, on the other hand, will provide tips or hints to users. It is an intuitive feature that will recommend links within the site. The WebSearch system is integrated to other search engines. It is a web tool, which can search for local sites as well as the Internet, online databases besides providing translations and other applications.

At the same time, the users can interfere and chat with the AINI chatterbots or Virtual Agent. Chatterbots use a form of artificial intelligence called natural-language processing to simulate conversations with users (Goh, Shahrin & Rajendraan, 2004). On the other hand, AINI also offers messaging, email and phone services to enable the users to interact with AINI in order to get more information. The components of AINI's system are shown in Figure 2.0.

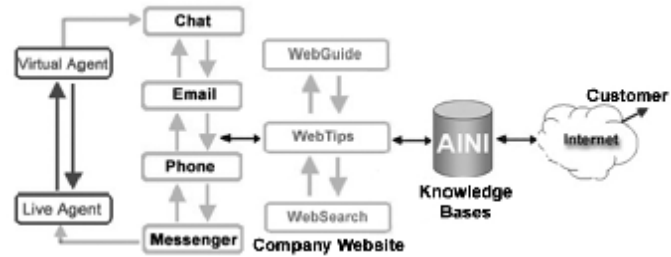

Figure 2.0: Components of AINI's System

### 4.1.2 PDA Chat

The idea of developing AINI into Personal Digital Assistance (PDA) is an interesting approach to having a more human and personalized interface between a computer and human. The PDA chat with AINI performs functions in web chats but with mobile compatibility. It is a prototype designed to blend mobile technology with natural language to help humans interact more naturally with mobile devices. Here, the implementation of PDA chat with the knowledge base will be designed using WiFi technology.

### 4.1.3 Mobile Chat

The mobile chat provides facilities such as SMS, MMS and GPRS services for users. All the services are integrated as one mobile chat component for providing latest alerts and information to users. Thus, the mobile chat is an alternative way where users can chat with AINI using SMS services. The SMS services will be on-subscribe mode, on-demand and on-alert news based services from the CCNet Portal. MMS is an SMS type service but with added image, voice, animation and many more features. Therefore, the users will be able to view images of viruses, bacteria, infected cells and also X-Ray images. Meanwhile, the GPRS technology provides mobile web browsing functionality for accessing the news on the CCNet Portal.

## 4.2 Back-End System (Server's Side)

Despite the tremendous growth in information technology, the storage of information and data is taking up great space in every computer's architecture system. Generally, every computer application system contains a database system. The idea of these approaches is to focus on the development of an intelligent database system known as 'knowledge base'. This 'knowledge base' is the brain of the CCNet portal, which contains the domain knowledge for crisis communication based on the medical field. Thus, all information in this knowledge base is extracted from the Knowledge Extraction Agent, which will be explained in detail in Section 4.2.2.

### 4.2.1 Knowledge Base

In this system, the knowledge base consists of AINI's common knowledge base, Expression Emotion Database, Customer knowledge base and AlertNews knowledge base to provide functionality of the portal. A detailed description of each component is provided in the following sections.

### 4.2.1.1 AINI's Knowledge Base

Based on the explanation above, AINI, which was developed by Goh Ong Sing (Goh & Chong, 2003; Goh & Fung, 2003; Goh, Mohd Ishak, Goh & Tee, 2003; Goh & Teoh, 2002), is written in pure Java. In this context, AIML is used to represent AINI's knowledge base. It is an XML specification for programming chat robots. The typical way of representing knowledge in an AIML file is:

```
<aiml>
<category>
<pattern>PATTERN</pattern>
<template>TEMPLATE</template>
</category>
</aiml>
```

The <aiml> tag indicates that this file describes the knowledge of a chatterbot. The <category> tag indicates an AIML category, the basic unit of the chatterbot's knowledge. The category has a <pattern> and <template>. This pattern represents the question and the template represents the chatterbot's answer (Goh & Teoh, 2002). Users' chat sessions with AINI in the website, which involve queering of any topic related to crisis communication will be processed here.

### 4.2.1.2 Expression Emotion Knowledge Base

The Expression Emotion Database, on the other hand, is embedded into the knowledge base to identify and classify emotions from the context of conversations with the software robot. Thus, AINI is able to interpret human speech and generate proper responses. The concept of communication between a human and the agent through AINI is depicted in Figure 3.0 (Goh, Goh & Tee, 2003).

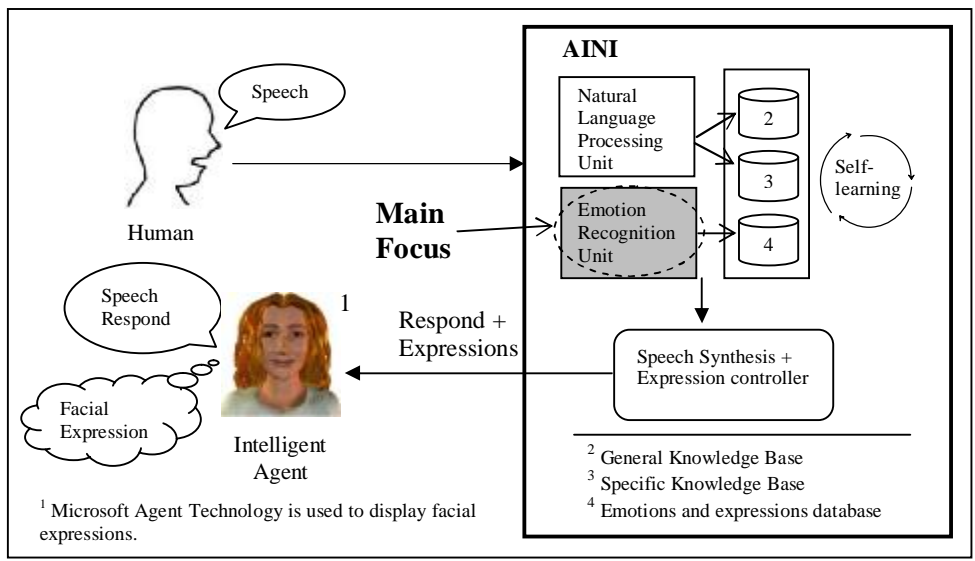

Figure 3.0: Communication between agent and human through AINI

From the diagram above, human speech will be passed to the natural processing unit in AINI for analysis and processing. Proper responses will be generated by the natural processing unit by extracting the knowledge stored in the database. The emotion recognition unit is responsible for identifying the emotion found in the speech and instructs the agent to display an appropriate facial expression. The interaction process can become more interesting and compelling, as the agent is able to show different expressions when it talks. For example, the agent will display a happy face to greet the user when the conversation starts; it will display a sad face when it hears something miserable; and it will show an angry face when the user says some obscene words.

In order to have a more attractive animated agent character's expression, an <agplay/> tag is created (Goh, 2001). This tag is an additional feature to the standard AIML tags and works alongside with an agplay Parser and Handler to make AINI read the chatterbot's answer, and play the expression supplied within the <agplay/> tag. Below is an example showing how the <agplay/> tag is used. This category is executed when the user greets the AINI by typing "HELLO". In return, the AINI will smile to the user and respond by saying "Hi there! How do you feel today?" (Goh & Teoh, 2002).

```
<category>
<pattern>HELLO</pattern>
<template>
        Hi there!  How do you feel today?
        <agplay anims="greet, pleased"/>
</template>
</category>
```

### 4.2.1.3 Customer's Knowledge Base

In fact, AINI's Knowledge Base actually depends on the configuration of the customer's knowledge base. The customer knowledge base stores the information on related topics that users request from the portal, PDA and mobile devices. Besides that, the customer knowledge base also keeps track of users' involvement with this website by the means of log analysis. This component not only keeps track of conversations between users and software robots, but also keeps track of the server log. Server logs are very useful in helping the developer understand the impact of the web site and in better positioning the web site to attract new and repeat visitors.

### 4.2.1.4 AlertNews Knowledge Base

The AlertNews knowledge base provides news and information to subscribed users who use mobile chat. As mentioned earlier, the mobile chat comprises of the SMS, MMS and GPRS technologies with special features such as on-subscribe, on-demand and on-alert news of AlertNews. There are three types of users of this system, which are the subscribed users, the non-subscribed users and the CCNet editorial. The subscribed users are those with the interest in crisis communication and are registered with the CCNet portal. They can post news, alert message etc by subscribing to the messaging list service of AlertNews. Meanwhile, non-subscribed users can only receive latest news besides having chat sessions with AINI. The CCNet editorial can also gain benefit by receiving latest AlertNews as a group member of the CCNet portal. Information is gained from the knowledge base of AINI with extra capability of providing latest information in the alert form under the AlertNews component. The AlertNews architecture is shown in Figure 4.0 below:

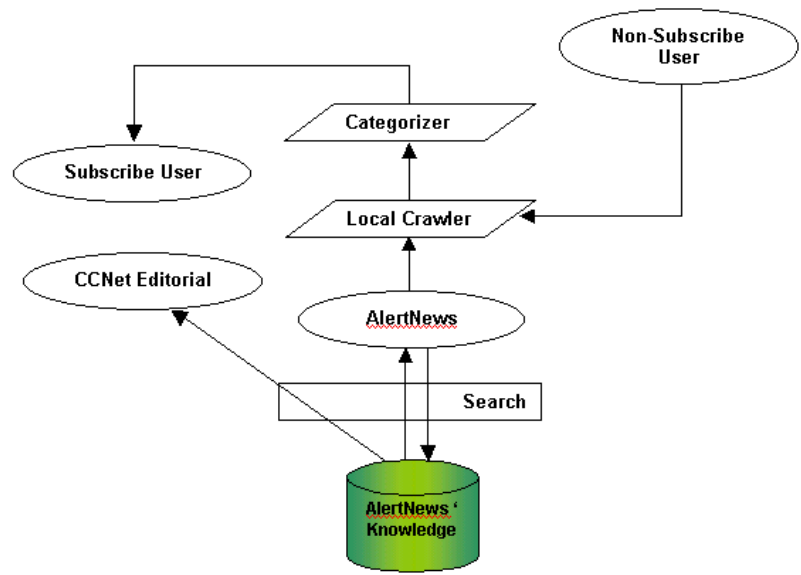

Figure 4.0: AlertNews Architecture

### 4.2.2 Knowledge Extraction Agent

The figure below shows the architecture of the CCNet Knowledge extraction agent. Four modules make up the agent with the crawler as the interface between the agent and the web. The crawler is like those used in conventional crawler-based search engines. The crawler resolves root domain names and follows subsequent links that is available on a page until a certain depth is defined by the user. These configurations are set in the crawler config database. For every page crawled, a copy is returned for further processing by the wrapper. The crawler's activities are logged in the crawler log database.

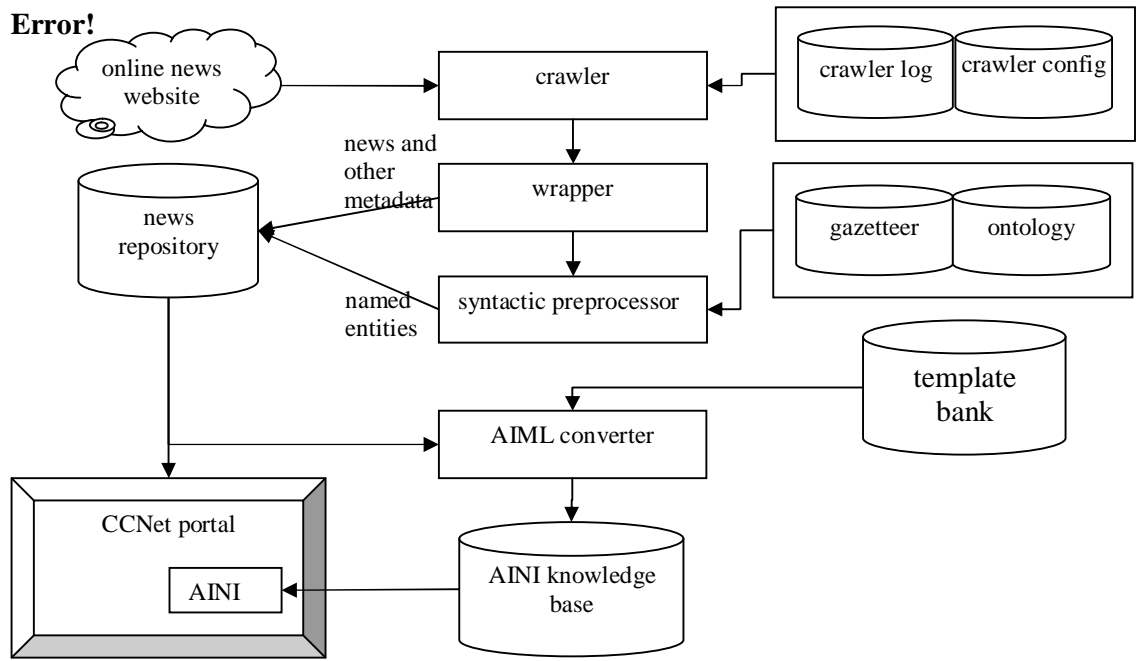

**Figure 4.1: CCNet knowledge extraction agent architecture**

Online news documents returned by the crawler are in the hypertext format and consist of a variety of unwanted characters (Figure 4.2). The wrapper prepares the raw news by separating the actual news content and other meta-information from hypertext characters. This process is also known as cleaning and the result is referred to as cleaned news (Figure 4.3). Information such as date of news, news title, news content and many more is extracted and stored in the CCNet news repository.

```
<html>
<head>
<title>New meningitis threat being contained by web of partnerships</title>
...
<p><font face="Times, Times New Roman, serif" size="3">
 8 APRIL 2004 | GENEVA -- A rare strain of meningitis, which re-emerged recently in Burkina
Faso…</font></p>
…
```

**Figure 4.2: Example of online news returned by crawler**

```
title: New meningitis threat being contained by web of partnerships
url: http://www.who.int/mediacentre/releases/2004/pr25/en/
date: 8 April 2004
content: A rare strain of meningitis, which re-emerged recently in Burkina Faso…
```

**Figure 4.3: Example of cleaned news returned by wrapper**

The syntactic preprocessor performs the task of identifying the dependencies among words. Based on the dependencies, grammatical relations (i.e. phrasal categories) like noun phrases, verb phrases and prepositional phrases are extracted using sentence parser for the English language like Link Grammar (Sleator & Temperley, 1993) and Minipar (Lin, 1998). The named entities in noun phrases are assigned with tags such as disease, location and person using the weighted gazetteer approach proposed by Wong & Goh (2004). A reference list known as gazetteer is used by the preprocessor. These tags enable the agent to identify what type of entity the corresponding noun phrases are and in which level and node do these entities belong to in the ontology. Pronouns are also resolved whenever necessary. The named entities that have been tagged are inserted into the corresponding entry in the news repository. For example, using the cleaned news in Figure 4.3, the syntactic preprocessor managed to identify two named entities namely meningitis and Burkina Faso. Using the gazetteer, the preprocessor will discover that meningitis is a type of disease and Burkina Faso is a country and tag them respectively using the ontology tag in the form *named_entity[ontology tag].*

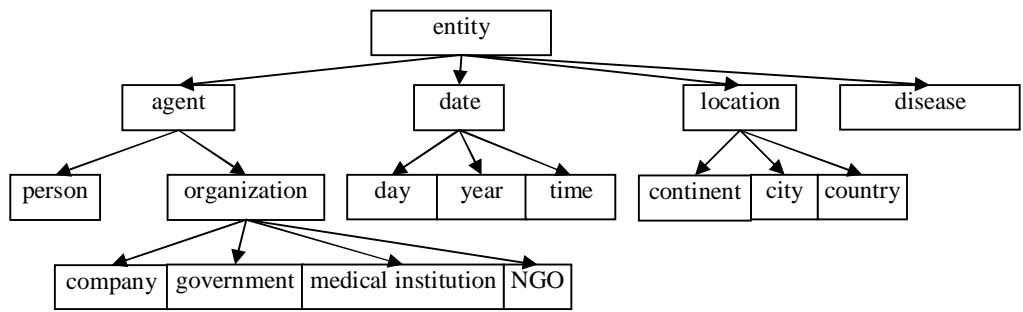

**Figure 4.4: CCNet disease ontology**

The information in the news repository is fed into two main components, namely the CCNet portal and the AIML converter. Information in the news repository is directly published to the CCNet portal without any further processing. Figure 4.5 below shows a sample entry in the news repository.

| url | date | title | content | named entity |
|-----|------|-------|---------|--------------|
| http://www.who.int/mediacentre/releases/2004/pr25/en/ | 8 April 2004 | New meningitis threat being contained by web of partnerships | A rare strain of meningitis, which re-emerged recently in Burkina Faso… | meningitis[disease] Burkina Faso[country] |

**Figure 4.5: A sample entry in the news repository**

The AIML converter uses the template bank to transform the news repository entries into AIML representation, which will be populated into AINI's knowledge base (Figure 4.6). The richer the template bank, the wider the scope of questions AINI will be able to handle.

```
<pattern> [wh-token corresponding to the ontology tag] _ [disease named_entity] _ </pattern>
<template> [first two lines of content]
<javascript>
window.open("[url]","","","");
</javascript>
</template>
```

**Figure 4.6: Sample template A from the template bank**

Substitutions will be made to the template using relevant values of each entry in the news repository. There are four fields in the template namely the wh-token corresponding to the ontology tag, first two lines of content, disease named entity and URL. The first and second requires some processing prior to replacement.

The ontology tag associated with each named entity is resolved to obtain the corresponding wh-token. Currently, the agent is capable of handling four types of wh-token: *where*, resolved from *location* named-entities, *when*, resolved from *date* named-entities, *who*, resolved from *agent* named-entities and what. The *what* token is resolvable from all ontological entities with additional tokens. For example, given the named entity *Burkina Faso* and its tag *country*, we can obtain the *where* token and *what* token with the *country* tag. This is possible because the question *where does meningitis...?* is similar to asking *what country does meningitis...?*(Figures 4.7 and 4.8).

```
<pattern> where _  meningitis _ </pattern>
<template> A rare strain of meningitis, which re-emerged recently in Burkina Faso…
<javascript>
window.open("http://www.who.int/mediacentre/releases/2004/pr25/en/","","","");
</javascript>
</template>
```

**Figure 4.7: Instance 1 of template A using *where* token**

```
<pattern> what country _  meningitis _ </pattern>
<template> A rare strain of meningitis, which re-emerged recently in Burkina Faso…
<javascript>
window.open("http://www.who.int/mediacentre/releases/2004/pr25/en/","","");
</javascript>
</template>
```

Figure 4.8: Instance 2 of template A using *what country* token

The second processing required prior to replacement is truncating the news content to the first two lines to be used in AINI's answers. The remaining news will be presented as part of a URL push.

The AIML converter follows precedence in converting named-entities and their ontology tag into AIML representation. All questions handled by AINI orbit the concept of disease and thus, all news content will surely contain *disease* named-entities. During conversion, priority will be given to entities other than disease. Only when a news does not contain any other entities (i.e. there are no information about *location*, *person* or *date*), then the converter will resolve the sole *disease* named-entity to the *what* token.

As an example, by referring to the sample entry in Figure 4.5, a *disease* named-entity and *country* named-entity exists. Due to the priority given to non-disease named-entity, the *country* named-entity is resolved to *where* token and *what country* token, creating two instances of the template. Then the remaining three fields are filled accordingly with the disease named entity (i.e. meningitis), the truncated news content (i.e. A rare strain of meningitis, which re-emerged recently in Burkina Faso…) and URL (i.e. http://www.who.int/mediacentre/releases/2004/pr25/en/).

Finally, these instances will be populated into AINI's knowledge base for learning and used by the AINI's chat interface in the CCNet portal for natural language question answering.

## 5.0  Conclusion

Nowadays, research on crisis communication cannot ignore the fact that the Internet has become an important source of communication in times of crisis, especially when they are global issues. Therefore, this research will prove that the Internet will create global virtual communities by using an intelligent agent software robot. This intelligent software robot helps by giving necessary and vital information needed by users during a crisis. It will also help in maintaining calm and order so that it will not trigger panic in the country. Users will have more trust in the information provided by the intelligent software robot because of its interactivity elements. Furthermore, the integrated TTS (Text-To-Speech) Technology and 3D human-like character or avatar in the system will able to generate speech and interact with the user in a humanlike manner. It also provides some news, advertisements, conversation logs and statistics in the system to benefit researchers in their effort for further enhancement of their systems. Besides that, new diseases are expected to hit world in the future due to mutation of bacteria and viruses. The output from this research can be used to tackle many health crises like the Bird Flu, AIDS, the mad cow disease, the Ebola virus, the Japanese Encephalitis (JE) virus, the *Nipah* virus and even old diseases that have re-emerged and become vital components of crisis communication.

**References**

Clark, H. H. (1999). How do real people communicate with virtual partners. In Brennan, S., Giboin, A. & Traum, D. (Eds.), *1999 AAAI Fall Symposium, Psychological Models of Communication in Collaborative Systems* (pp. 43-47), CA: AAI Press.

Goh O. S., Shahrin Sahib &  Rajendraan Elangsegaran. (2004). An Intelligent Virtual Financial Advisor System (IVFAS). In M.H. Hamza (Ed.), *2nd IASTED International Conference on Neural Network and Computational Intelligence (NCI 2004)* (pp. 146-151). Switzerland: ACTA Press.

Goh O. S. & Chong K. E. (2003). Design Methodologies for Humanoid WebCall Center. In Batovski, D.A. (Ed.), *2003 International Conference on Information and Communication Technology (ICT 2003)* (pp. 344-349). Thailand: Assumption University

Goh, O.S. & Fung, C.C. (2003). Intelligent Agent Technology in E-Commerce. *Intelligent Data Engineering and Automated Learning, LNCS 2690*, 10-17.

Goh O. S., Goh, M. & Tee, C. (2003). Emotion Recognition in Intelligent Agent [CD-ROM].

Goh, O. S., Mohamad Ishak Desa, Goh K. O. & Tee, C. (2003). An Intelligent Web 3D for E-Commerce. In M. Mohammadian (Ed.), *International Conference on Intelligent Agents, Web Technology and Internet Commerce - AWTIC'2003* (Book of Abstracts) (pp. 42-43). Australia: University of Canberra.

Goh, O. S. & Teoh K. K. (2002). Intelligent Virtual Doctor System. Paper presented at 2nd IEE Seminar on Appropriate Medical Technology for Developing Countries, London, United Kingdom. Available http://www.iee.org/Oncomms/pn/healthtech/06Feb.cfm,

Goh, O. S. (2001). <agplay> Custom AIML Tag. Available: http://fist.mmu.edu.my/msc/doc/agplayTag.html

Halavais, A. (2000). National borders on the World Wide Web. *New Media & Society.* 2(1),7-28.

Life Services Network. (1996). *Crisis Communication Handbook*. United States: Public Relation Task Force.

Lin, D. (1998). *Dependency-Based Evaluation Of MINIPAR*. Paper presented at the 1st International Conference on Language Resources and Evaluation, Spain.

Sleator, D. & Temperley, D. (1993). *Parsing English With A Link Grammar*. Paper presented at the 3rd International Workshop on Parsing Technologies, Prague, Czech Republic.

Smith, C. (2003, April - June). Crisis Communication in the Internet Age. *National Capital Chapter of the Public Relations Society of America.* Available: http://www.prsa-ncc.org/news/newsLetterarticle.asp?insertPage=NL03spring9.htm

US and Canada warn websites about claims (2003, May 11), *The Star*, pp. 11. Available: http://www.thestar.com.my/news/archives/story.asp?ppath=\2003\5\11&file=/2003/5/11/world/ubatbest&sec=world

WHO Communicable Disease Surveillance & Response (CSR). (2004). The Global Outbreak Alert and Response Network. Available: http://www.who.int/csr/sarscountry/2003_05_12/en/
Wong, W. & Goh, O. S. (2004). *Syntax Preprocessing in Cyberlaw Web Knowledge Base Construction*. To be presented at the International Conference on Computational Intelligence for Modelling, Control and Automation – CIMCA2004, 12-14 July 2004 Gold Coast, Australia.

World Health Organization. (2003). Cumulative number of reported probable cases. Available: http://www.who.int/csr/sarscountry/2003_05_12/en/